Title

# Time Dependent Quadratic Hamiltonians, SU(1,1), SU(2), SU(2,1) and SU(3).

Author

Paul Croxson

Address

Formally Phd student of Maths Dept, King's College London, Strand, London WC2R 2LS, UK now seeking a position.
emails; < vincecroxson@hotmail.com >, < pcroxson@mth.kcl.ac.uk >.

Date  April 2006

## Abstract

*The properties of SU(1,1) SU(2), SU(2,1) and SU(3) have often been used in quantum optics. In this paper we demonstrate* the *use of these symmetries. The group properties of SU(1,1) SU(2), and SU(2,1) are used to find the transition probabilities of various time dependent quadratic Hamiltonians. We consider Hamiltonians representing the frequency converter, parametric amplifier and raman scattering. These Hamiltonians are used to describe optical coupling in nonlinear crystals.*

## Introduction

The main point of this paper is to find transition probabilities, for some multi - mode quadratic time dependent Hamiltonians taken from quantum optics. Many of these Hamiltonians take the form of a linear combination of the generators of a Lie Algebra with time dependent coefficients.

The methods adopted in this work express the time-dependent unitary transformation for the evolution of a time-dependent system. which is a commonly recurring problem. This is done in group theoretical terms
Time dependent Hamiltonians occur in many areas of quantum optics, some examples of this are the Paul trap [1], ultra - short laser pulses [2]. optical coupling in non - linear crystals, creating squeezed and entangled states,[3 - 7].

We wish to apply the matrix disentangling method to some specific multi - mode time dependent quadratic Hamiltonians, consisting of the generators of a Lie algebra whose coefficients may be time dependent.

Using this disentangling method we will be able to find the transition probabilities for these Hamiltonians. Transition probabilities show how the modes fluctuate. These mode fluctuations occur when Laser light passes through non - linear crystals
We consider two and three mode quadratic Hamiltonians.
The Hamiltonians are described by a combination of Lie group generators, with

time - dependent coefficients.

In fact, the use of matrix representation is usually only applied when the Hamiltonians under consideration are time independent. We show how to extend this method to some time dependent Hamiltonians, important in the field of quantum optics. We will use the group properties of SU(2), SU(1,1), and SU(2,1) to accomplish this task. The three Hamiltonians we consider from quantum optics are the off - resonant frequency converter, the parametric amplifier, (both two mode) and Raman scattering (three mode).

These Hamiltonians have been considered by many authors [3 - 5] and references therein. Work on the resonant case is found in [5]. Other quadratic time dependent Hamiltonians have been considered by various authors [6 - 9]. A matrix disentangling method has been proposed by Gilmore [10] and used by Santiago [11], Arecchi [12] to find the rotation operator.

Santiago [11] has presented formulas for disentanglement (decomposition) using the matrix properties of SU(2) and SU(1,1). These same formulas have been derived by Ban[13] using parametric differentiation, in fact this author shows how to apply these and other formulas to the Mach - Zehnder interferometer, an important device in optical experiments. Other authors have used the methods of parametric differentiation [14]

One problem with this matrix method is that it can only be applied to Hamiltonians with no explicit time dependency. Zahler and Aryeh [15] have overcome this problem for a single mode, off - resonant, Hamiltonian with a specific time dependency.

Cheng and Fung [16] have used group properties of SU(2) to find the evolution operator for the time dependent mass - varying harmonic oscillator. The decomposition of symmetry groups of higher order has been considered by [17 - 18]. They use Laplace transforms to accomplish this task. We illustrate their work and also apply the techniques adopted in this paper to the decomposition of SU(3).

We will consider two mode and three mode time dependent, off - resonant Hamiltonians. In this paper the proposed procedure of finding the transition probability is purely algebraic, in that there is no need to solve any differential equations.

Other disentangling methods are considered by [19]. Their approach is applicable to problems involving time dependent quadratic Hamiltonians and leads to sets of coupled differential equations. As Dattoli [19] states, the method they employ can lead to sets of coupled differential equations that,
"do not appear easily solvable even in the case in which all the couplings are constant. "
This is particularly relevent to three mode Hamiltonians .The approach we adopt here is not as widely applicable, but leads to explicit expressions for the transition probabilities for some particular time dependent couplings.

**Section 1 Review**

To put the above task in context, we consider the work of Dattoli[19] and others who have used a disentangling method based on the algebraic properties of the Lie algebra. We outline the method they use below.
This method leads to sets of coupled differential equations, which need to be solved if the desired transition probabilities are to be found.
Many have adopted a group approach to problems in quantum optics [20]
We note that in [20] the techniques used by the authors are similar to those used by [19].
Consider the Hamiltonian represented by,

$$H = \sum_{j=1}^{m} \omega_j(t) L_j ,  \tag{1.1}$$

where the operators, $L_j$, form a Lie algebra given by,

$$[L_j, L_k] = \sum_{r=1}^{n} \beta_{jk}^{r} L_r  \tag{1.2}$$

and $\omega_j(t)$, may be explicitly time dependent. Here, '$n$', is the dimension of the Lie algebra. The solution to Schrödinger's equation,

$$ih\frac{\partial U}{\partial t} = HU, \qquad (1.3)$$

is then the evolution operator, $U(t)$, written in the disentangled form as,

$$U(t) = \prod_{j=1}^{n} \exp(g_j(t) L_j) \qquad (1.4)$$

where the functions, $g_j(t)$ are given by the set of first order coupled differential equations,

$$\begin{pmatrix} \omega_1(t) \\ \omega_2(t) \\ \cdot \\ \cdot \\ \omega_n(t) \end{pmatrix} = \begin{pmatrix} \xi_{1,1} & \cdot & \cdot & \cdot & \xi_{1,n} \\ & \cdot & \cdot & \cdot & \cdot \\ \cdot & \cdot & \cdot & \cdot & \cdot \\ \cdot & \cdot & \cdot & \cdot & \cdot \\ \xi_{n,1} & \cdot & \cdot & \cdot & \xi_{n,n} \end{pmatrix} \begin{pmatrix} \dot{g}_1(t) \\ \dot{g}_2(t) \\ \cdot \\ \cdot \\ \dot{g}_n(t) \end{pmatrix} , \qquad (1.5)$$

the, $\xi_{i,j}$ depend on the algebra structure constants, $\beta_{jk}^{r}$, and the overdot, $\cdot$, means time derivative. The disentangled form of the evolution operator was then used by Dattoli to find the transition probabilities for a specific example from quantum optics, a Hamiltonian with SU(2) symmetry.

We will consider a set of Hamiltonians, from quantum optics with this symmetry, as well as SU(1,1) and SU(2,1), which are explicitly time dependent. We find their corresponding transition probabilities between specific number states but base our analysis on a representative matrix algebra. These transition probabilities have not previously been evaluated.

## Section 2 Frequency Converter, SU(2)

The first Hamiltonian we consider is the off - resonant two mode frequency converter. This has a SU(2) symmetry [3.4] and its explicit time dependence is shown below,

$$H_c = h\omega_a a^{\dagger}a + h\omega_b b^{\dagger}b + h k(a^{\dagger}b\, e^{-i(\omega_a - \omega_b)t + i\,t} + h.c.) , \qquad (2.1a)$$

with evolution operator, $U_C(t)$, given by the evolution equation,

$$H_c U_C(t) = ih\frac{\partial}{\partial t} U_C(t). \qquad (2.1b)$$

To proceed with finding its transition probability, we need to find an expression for the evolution operator $U_C(t)$ above. This is achieved by removing the explicit time dependency from this Hamiltonian, $H_c$, (2.1a). We transform Hamiltonian $H_c$, into a time independent interaction Hamiltonian.

Choosing interaction operators, $a_I$, $b_I$ to be,

$$a_I = exp\left(\frac{iH_{I0}t}{h}\right) a \, exp\left(\frac{-iH_{I0}t}{h}\right) = a \, e^{i\omega_a t - i\delta_a t}, \quad (2.2a)$$

where, $\quad H_{I0} = h(\omega_a - \delta_a)a^\dagger a + h(w_b + \delta_b)b^\dagger b,$ (2.2b)

with the condition, $\quad \delta = \delta_a + \delta_b$.

Using a similar transformation to (2.2a) we get
$$b_I = b \, e^{i\omega_b t + i\delta_b t}. \quad (2.3)$$

We then get interaction Hamiltonian $H_{II}$,[3,4]

$$H_{II} = h\delta_a a_I^\dagger a_I - h\delta_b b_I^\dagger b_I + hk(a_I^\dagger b_I + h.c.).$$
(2.4)

Then the transition matrix elements, from an initial number state, $|n>$, to a final number state, $|m>$ say, for this Hamiltonian (2.1a) takes the form (appendix A),

$$S_{m,n}(t) = <m | U_C(t) | n> \; = <m | U_{I0}(t)U_{II}(t) | n>, \quad (2.5)$$
where,
$$U_{I0}(t) = exp(-iH_{I0}t/h) \; , \text{ and } \; U_{II}(t) = exp(-iH_{II}t/h) \quad (2.6)$$

In (2.5) the bra $<m|$ labels the final number state, $<m_a | <m_b |$, where $<m_a|$ is the final number state for mode 'a', $<m_b|$ is the final number state for mode 'b'. Similarly, $|n>$, is the initial number state $|n_a> |n_b>$.

### 2.1 Matrix representation
We will now see how we can use a faithful matrix representation of the creation(annihilation) operators to find the transition probabilities.
Using (2.5, 2.6) and the properties of these operators acting on the number state, $|n>$,

$$a^\dagger | n_a> = \sqrt{n_a + 1} | n_a + 1>, \quad a | n_a> = \sqrt{n_a} | n_a - 1> , \quad (2.7)$$

we get, after some operator algebra,[21,25] the transition probability,

$$|S_{m,n}(t)|^2 = |<m| \exp\ -it\{(\omega_a + \omega_b)\frac{(a_I^\dagger a_I - b_I^\dagger b_I)}{2} + k(a_I^\dagger b_I + h.c.)\} |n_a>|n_b>|^2. \quad (2.8)$$

The operators which occur in (2.8) are the generators of su(2) Lie algebra, $J_3, J_+$, and $J_-$, where we can choose,

$$J_3 \equiv \frac{(a_I^\dagger a_I - b_I^\dagger b_I)}{2}, \quad J_+ \equiv a_I^\dagger b_I, \quad \text{and}\ J_- \equiv a_I b_I^\dagger, \quad (2.9)$$

and have commutation relations,

$$[J_3, J_\pm] = \pm J_\pm, \quad [J_+, J_-] = 2J_3. \quad (2.10)$$

A particular faithful matrix representation of this is,

$$J_- = \begin{pmatrix} 0 & 0 \\ 1 & 0 \end{pmatrix}, \quad J_+ = \begin{pmatrix} 0 & 1 \\ 0 & 0 \end{pmatrix}, \quad \& \quad J_3 = \frac{1}{2}\begin{pmatrix} 1 & 0 \\ 0 & -1 \end{pmatrix}. \quad (2.11)$$

We now use the results of [11,13]. Tthis will enable us to write the distanglement formulas for SU(2). From the formula given in [11,13] and above we can write equ(2.8) as,

$$|S_{m,n}(t)|^2 = |<m| \exp[X_+ J_+]\exp[(lnX_3)J_3]\exp[X_- J_-] |n_a>|n_b>|^2, \quad (2.12)$$

In (2.12) the variables $X_-$, $X_+$, and $X_3$, are,

$$X_\pm = \frac{W_\pm \sinh f}{2f \cosh f - W_3 \sinh f}, \quad X_3 = [\cosh f - (W_3/2f)\sinh f]^{-2}, \quad f^2 = \tfrac{1}{4}W_3^2 + W_- W_+. \quad (2.13)$$

with,

$$W_3 \equiv -it(\omega_a + \omega_b) = -it\omega, \quad W_+ \equiv -ikt, \quad W_- \equiv -ikt. \quad (2.14)$$

To demonstrate these formulas we consider an example.

## 2.2 An Example

We calculate the transition probability in going from an initial vacuum state '|0>' in mode 'b' to final vacuum state '|0>' in mode 'a', i.e.

$$|n> = |N>_a |0>_b, \quad |m> = |0>_a |N>_b. \quad (2.15)$$

Then after some algebra, we find that, from (2.8, 2.12,& 2.13),

$$|S_{m,n}(t)|^2 = \left[\frac{k \sin\sqrt{k^2 + (\omega/2)^2}\ t}{2\sqrt{k^2 + (\omega/2)^2}}\right]^{2N}. \quad (2.16)$$

We see in (2.16) the oscillatory nature of the transition probability. In this example we have chosen an initial vacuum state for one of the modes.

## Section 3 Parametric Amplifier, SU(1,1)

We can adopt a similar proceedure to that in section (2) above. We wish to find the transition probability for the off resonant parametric amplifier. It has a SU(1,1) symmetry and its Hamiltonian is represented by [3, 4],

$$H_a = h\omega_a a^\dagger a + h\omega_b b^\dagger b + hk(a^\dagger b^\dagger e^{-i(\omega_a + \omega_b)t + i\Delta t} + h.c.). \tag{3.1a}$$

with evolution operator, $U_A(t)$, given by the solution to the evolution equation

$$H_a U_A(t) = ih\frac{\partial}{\partial t} U_A(t). \tag{3.1b}$$

The explicit time dependency is removed from the Hamiltonian, $H_a$, by choosing new operators to be,

$$a_2 = exp\left(\frac{iH_{20}t}{h}\right) a\, exp\left(-\frac{iH_{20}t}{h}\right) = ae^{i\omega_a t - i\Delta_a t}, \text{ similarly, } b_2 = be^{i\omega_b t - i\Delta_b t}, \tag{3.2}$$

where the off resonance condition is characterised by the detuning parameter,

$$\Delta = \Delta_a + \Delta_b, \tag{3.3}$$

and,

$$H_{20} = h(\omega_a - \Delta_a)a^\dagger a + h(\omega_b - \Delta_b)b^\dagger b, \tag{3.4}$$

For this parametric amplifier, with these new operator variables, '$a_2$' and ,'$b_2$', the explicit time dependence is removed and the Hamiltonian is now,[4]

$$H_{2I} = h\Delta_a a_2^\dagger a_2 + h\Delta_b b_2^\dagger b_2 + h k(a_2^\dagger b_2^\dagger + h.c.), \tag{3.5}$$

The transition matrix elements for the parametric amplifier, $S_{m,n}(t)$, then take the form (see appendix A),

$$S_{m,n}(t) = <m|U_A(t)|n> = <m|U_{20}(t)U_{2I}(t)|n> \tag{3.6}$$

where,

$$U_{20}(t) = exp(-iH_{20}t/h). \qquad U_{2I}(t) = exp(-iH_{2I}t/h).$$

and the bra, $<m|$, and ket, $|n>$, are final and initial number states, $<m_a|<m_b|$ and $|n_b>|n_a>$, respectively.

### 3.1 Matrix Representation

Letting the initial number state be $|n> = |n_a>|n_b>$, $|m>$ being the final number state as before, after some algebra the transition probability becomes,

$$|S_{m,n}(t)|^2 = |<m| \exp\ -it\{(\omega_a + \omega_b)\frac{(a_I^\dagger a_I + b_I^\dagger b_I)}{2} + k(a_I^\dagger b_I^\dagger + h.c.)\}\ |n_a>|n_b>|^2$$

(3.7)

Similar to the case of the frequency converter, we identify the operators, of the su(1,1) Lie algebra as

$$\frac{(a_I^\dagger a_I + b_I^\dagger b_I)}{2},\ a_I^\dagger b_I^\dagger,\ \text{and},\ a_I b_I .$$

(3.8a)

If we choose to define,

$$K_3 \equiv \frac{(a_I^\dagger a_I + b_I^\dagger b_I)}{2},\quad K_+ \equiv a_I^\dagger b_I^\dagger,\ \text{and}\ K_- \equiv a_I b_I.$$

(3.8b)

then we have commutation relations,

$$[K_3, K_\pm] = \pm K_\pm,\ [K_+, K_-] = -2K_3.$$

(3.9a)

One particular matrix representation of this su(1,1) algebra is,

$$K_- = \begin{pmatrix} 0 & 0 \\ -1 & 0 \end{pmatrix},\ K_+ = \begin{pmatrix} 0 & 1 \\ 0 & 0 \end{pmatrix},\ K_3 = \frac{1}{2}\begin{pmatrix} 1 & 0 \\ 0 & -1 \end{pmatrix}.$$

(3.9b)

We then write,

$$|<m| \exp\ -it\{(\omega_a + \omega_b)\frac{(a_I^\dagger a_I + b_I^\dagger b_I)}{2} + k(a_I^\dagger b_I^\dagger + h.c.)\}\ |n_a>|n_b>|^2$$
$$= |\exp(Y_+ K_+)\exp((\ln Y_3)K_3)\exp(Y_- K_-)|^2 .$$

(3.10)

The variables, $Y_+$, $Y_-$, and $Y_3$ are given in [11,13] are,

$$Y_\pm = \frac{V_\pm \sinh F}{2 F \cosh F - V_3 \sinh F},\quad Y_3 = [\cosh F - (V_3/2F)\sinh F]^{-2}.\ F^2 = \tfrac{1}{4} V_3^2 - V_- V_+.$$

(3.11)

To demonstrate these formulas we consider an example.

*3.2 An Example*

In the formula (3.7), we consider the initial number states, $|n_a>$, $|n_b>$, to be the vacuum state, $|0>$, let, $|n> = |0_a>|0_b> = |0>$, and the final number states be, $|m_a> = |m_b> = |1>$, let $|m> = |1_a>|1_b> = |1>$. This will enable us to examine conditions for either growth or oscillation for the parametruic amplifier. With these photon number states, the transition probability is,

$$|S_{1,0}(t)|^2 = \left| \frac{V_+ \sinh F}{2F \cosh F - V_3 \sinh F} \right|^2, \qquad (3.12a)$$

where,

$$F = \sqrt{\tfrac{1}{4} V_3^2 - V_- V_+} = it\sqrt{(-k^2 + (\ /2)^2)}, \qquad (3.12b)$$

with,

$$V_- = V_+ = -ikt, \quad V_3 = -it(\ _a + \ _b) = -it\ . \qquad (3.13)$$

We note, in (3.12), that the condition for growth or oscillation is given by whether ,

$$k^2 > (\ /2)^2, \quad \text{or} \quad k^2 < (\ /2)^2, \qquad (3.14)$$

a weak coupling constant ,'k' , leads to an oscillation of the modes. In this case we have chosen an initial vacuum state for both modes.

**Section 4 Raman Scattering, SU(2,1)**

In this section we discuss the technique, outlined above, applied to a linear combination of SU(2,1) generators. These generators will have some specific time dependent coupling coefficients. Unlike, SU(2), SU(1,1) and SU(3),there does not appear to be an example of the decomposition of SU(2,1) in the literature, so we will show how this is acheived.

In appendix B we discuss SU(3). The well known SU(3), among other things,has applications in elementary particle physics and Schiff [22] applies it to a three mode case, Leach [23] uses it to find invariants of quadratic Hamiltonians. Rowe, Sanders and de Guise construct a basis for irreducible representations for SU(3),[24].

As noted by Dattoli in [19] the problem of finding the transition matrix elements for the SU(3) case leads to a set of three coupled differential equations that, ' do not appear easily solvable even in the case in which all the couplings are constant '.
We will find the transition probability between some specific number states for a Hamiltonian with time dependent coupling coefficients, i.e
the Hamiltonian for Raman scattering [3], [4],

$$H_r = h\omega_v a_v^\dagger a_v + h\omega_s a_s^\dagger a_s + h\omega_a a_a^\dagger a_a - hg_s(a_s^\dagger a_v^\dagger e^{-i(\omega_s + \omega_v)t + ik_s t} + h.c.) - hg_a(a_a^\dagger a_v e^{-i(\omega_a - \omega_v)t + ik_a t} + h.c.). \qquad (4.1)$$

This has an evolution operator, $U_r(t)$, given by the solution to evolution equation,

$$H_r U_r(t) = ih \frac{\partial}{\partial t} U_r(t). \qquad (4.2)$$

We will find this evolution operator $U_r(t)$, as we did in previous sections.
First we make transformations to new operators , given by,

$$b_s = e^{iH_{30}t/h} a_s e^{-iH_{30}t/h} = a_s e^{i\omega_s t - i k_s t}, \quad b_a = e^{iH_{30}t/h} a_a e^{-iH_{30}t/h} = a_a e^{i\omega_a t - i k_a t},$$
$$b_v = e^{iH_{30}t/h} a_v e^{-iH_{30}t/h} = a_v e^{i\omega_v t}. \quad (4.3a)$$

where,

$$H_{30} = h\omega_v a_v^\dagger a_v + h(\omega_s - k_s) a_s^\dagger a_s + h(\omega_a - k_a) a_a^\dagger a_a, \quad (4.3b)$$

with evolution operator,

$$U_{30}(t) = \exp((-iH_{30}t/h). \quad (4.3c)$$

For Raman scattering, with these new operator variables, '$b_v$' $b_a$', and '$b_s$', the explicit time dependence is removed and the Hamiltonian (4.1) is now,

$$H_{3I} = hk_s b_s^\dagger b_s + hk_a b_a^\dagger b_a - hg_s(b_s^\dagger b_v^\dagger + h.c.) - hg_a(b_a^\dagger b_v + h.c.), \quad (4.4)$$

with evolution operator,

$$U_{3I}(t) = \exp(-iH_{3I}t/h). \quad (4.5b)$$

Finally, we write the transition matrix elements for Raman scattering, $S_{m,n}(t)$, as [21,p58] ( appendix A),

$$S_{m,n}(t) = <m | U_r(t) | n> = <m | U_{30}(t) U_{3I}(t) | n> \quad (4.6)$$

*4.1 Matrix Representation*

We now utilise a matrix representation. This will help us evaluate (4.6).
Consider the Lie algebra su(2,1) written in terms of creation (annhilation) number operators,

$$A = b_a b_v^\dagger, B = b_a^\dagger b_v, C = (1/2)(b_v b_v^\dagger - b_a b_a^\dagger), \quad (4.7a)$$
$$D = b_s^\dagger b_v^\dagger, E = b_s b_v, F = (1/2)(b_v b_v^\dagger + b_s^\dagger b_s), \quad (4.7b)$$
$$G = b_a^\dagger b_s^\dagger, J = b_a b_s, K = (1/2)(b_a^\dagger b_a + b_s b_s^\dagger), \quad (4.7c)$$
with commutators, $[A,B] = 2C$, $[D,E] = -2F$, $[G,J] = -2K$, etc and $F - C = K$. $\quad (4.7d)$

As noted in [19] this has a group structure of the type,
 SU(1,1)   SU(1,1)   SU(2).

In (4.7) we note that, A,B, are the operators occuring in. the frequency converter, while D,E,G,J, are operators occuring in the parametric amplifier. The algebra (4.7d) is su(2,1)
In appendix B we consider operators all occuring in the frequency converter, which has algebra su(3).
A faithful matrix representation of this su(2,1) algebra is,

$$A \begin{pmatrix} 0 & 1 & 0 \\ 0 & 0 & 0 \\ 0 & 0 & 0 \end{pmatrix}; B \begin{pmatrix} 0 & 0 & 0 \\ 1 & 0 & 0 \\ 0 & 0 & 0 \end{pmatrix}; C\ (1/2)\begin{pmatrix} 1 & 0 & 0 \\ 0 & -1 & 0 \\ 0 & 0 & 0 \end{pmatrix}, E \begin{pmatrix} 0 & 0 & 0 \\ 0 & 0 & 1 \\ 0 & 0 & 0 \end{pmatrix};$$

$$D \begin{pmatrix} 0 & 0 & 0 \\ 0 & 0 & 0 \\ 0 & -1 & 0 \end{pmatrix}; F\ (1/2)\begin{pmatrix} 0 & 0 & 0 \\ 0 & -1 & 0 \\ 0 & 0 & 1 \end{pmatrix}, J \begin{pmatrix} 0 & 0 & 1 \\ 0 & 0 & 0 \\ 0 & 0 & 0 \end{pmatrix}; G \begin{pmatrix} 0 & 0 & 0 \\ 0 & 0 & 0 \\ -1 & 0 & 0 \end{pmatrix};$$

$$K \quad (1/2) \begin{matrix} -1 & 0 & 0 \\ 0 & 0 & 0 \\ 0 & 0 & 1 \end{matrix} , \qquad \text{hence, as in (4.7d), } [A,B] = 2C, \text{etc.} \tag{4.8}$$

We now find the transition probability (4.6),
$$| S_{m,n}(t) |^2 = | < m | U_{30}(t) U_{3I}(t) | n > |^2, \tag{4.9}$$
and introduce the operator,
$$I_I = h\omega(b_v^\dagger b_v + b_a^\dagger b_a - b_s^\dagger b_s), \tag{4.10}$$

This operator, "$I_I$", will enable us to write the transition probability, $| S_{m,n}(t) |^2$, in terms of the generators of SU(2,1).
Note that the commutator, $[H_{3I}, I_I] = 0$, $H_{3I}$ is given by (4.4).and '$\omega$' ,is a free constant to be determined.
We rewrite the transition probability as,

$$| S_{m,n}(t) |^2 = | < m | U_{30}(t) exp(-it\, I_I /h) exp(it\, I_I /h) U_{3I}(t) | n > |^2. \tag{4.11a}$$

As it can be shown that, $| < m | U_{30}(t) exp(-it\, I_I /h) | n > |^2 = 1,$ (4.11b)
we can write (4.11a) as,

$$| S_{m,n}(t) |^2 = | < m | exp(it\, I_I /q) U_{3I}(t) | n > |^2. \tag{4.12}$$

Choosing ,'$\omega$' in, '$I_I$', to be, $\omega = \dfrac{k_s - k_a}{3}$, leads to,

$$| S_{m,n}(t) |^2 = \left| < m | exp\{ it[(\dfrac{-k_s - 2k_a}{3})K + (\dfrac{k_a - k_s}{3})F - g_a(A+B) - g_s(E+D)]\} | n > \right|^2 \tag{4.13}$$

where we define for convenience, the Hermitian operator,

$$H \quad h\{(\dfrac{-k_s - 2k_a}{3})K + (\dfrac{-k_s + k_a}{3})F - g_a(A+B) - g_s(E+D)]\}. \tag{4.14}$$

Then the operator, $exp\left(\dfrac{it}{h} H\right)$, in equ(4.13) is anti-Hermitian.

To proceed, we write (4.13), in a decomposed symmetric form, with time dependent functions $f_i(t)$, as ,

$$| S_{m,n}(t) |^2 = | < m | exp(f_6 D) exp(f_4 G) exp(f_8 A) exp(f_0 C) exp(f_1 K) exp(f_2 F)$$
$$\times exp(f_7 B) exp(f_3 J) exp(f_5 E) | n > |^2, \tag{4.15}$$

the generators could have been written in another order and this would change the functions $f_i(t)$.

We could set, '$f_0$' above to zero since there are only eight independent generators. The , '$f_i$' can be determined by using the matrix representation of the generators and equating the two matrix forms (4.13) and (4.15). This process, involving lengthy matrix algebra, will give '$f_i$' in terms of the structure constants , the coupling constants, $g_s$, $g_a$ and the detuning parameters, $k_s$, $k_a$ .
We show here explicitly $f_4$ and $f_6$. These will be needed in a later example, details of this calculation are given in [25] and techniques used can be found in [26]. We get,

$$f_6 = -\frac{(u_1 w_1')e^{it\lambda_1} + (u_2 w_2')e^{it\lambda_2} + (u_3 w_3')e^{it\lambda_3}}{(w_1 w_1')e^{it\lambda_1} + (w_2 w_2')e^{it\lambda_2} + (w_3 w_3')e^{it\lambda_3}}, \tag{4.16a}$$

$$f_4 = \frac{(v_1 w_1')e^{it\lambda_1} + (v_2 w_2')e^{it\lambda_2} + (v_3 w_3')e^{it\lambda_3}}{(w_1 w_1')e^{it\lambda_1} + (w_2 w_2')e^{it\lambda_2} + (w_3 w_3')e^{it\lambda_3}}, \tag{4.16b}$$

where , $v_i$ , $u_i$ , and $w_i$ are the elements of the column eigenvectors of matrix (4.14) and, $v_i'$ , $u_i'$, and $w_i'$ are the elements of the row eigenvectors of matrix, (for more details see [25, appendix 2a]
Since the non-symmetric representative matrix of Hamiltonian, '$H$', (4.14) is,

$$H = h \begin{pmatrix} \frac{k_s + 2k_a}{6} & -g_a & 0 \\ -g_a & \frac{k_a - k_s}{6} & -g_s \\ 0 & g_s & \frac{-2k_s - k_a}{6} \end{pmatrix}, \tag{4.17}$$

we therefore have the elements of the column eigenvectors being,

$$v_i = [(-\lambda_i + \frac{k_a - k_s}{6})(-\lambda_i + \frac{2k_s + k_a}{6}) + g_s^2]/N_i, \quad u_i = [-g_a(-\lambda_i + \frac{2k_s + k_a}{6})]/N_i ,$$
$$w_i = (-g_s g_a)/N_i, \tag{4.18}$$

and normalising factor,' $N_i$ ', is given by, $v_i v_i' + u_i u_i' + w_i w_i' = 1$.

The eigenvalues, $\lambda_i$ will be given by the roots of the characteristic cubic equation of matrix (4.17).

*A Specific Example*
We wish to calculate the transition probability $|S_{m,0}(t)|^2$, going from an initial vacuum state $|n> = |0_a, 0_s, 0_v,>$, to a final number state, $|m> = |m_a, m_s, m_v>$ .
Using (4.15) we get,

$$|S_{m,0}(t)|^2 = |<m_v, m_s, m_a | exp(f_6 D)exp(f_4 G)exp(f_8 A)exp(f_0 C)exp(f_1 K)exp(f_2 F)$$
$$\times exp(f_7 B)exp(f_3 J)exp(f_5 E) | 0,0,0 >|^2. \tag{4.19}$$

Using the properties of the creation(annihilation) number operators,

$$a_a^\dagger | n_a > = \sqrt{n_a + 1} | n_a + 1 >, \quad a_a | n_a > = \sqrt{n_a} | n_a - 1 >, \quad (4.20)$$

we get, using (4.7),

$$exp(f_5 E) | 0 > = |0>, \, exp(f_7 B) | 0 > = |0>, \quad exp(f_3 J) | 0 > = |0>$$
$$exp(f_8 A) | 0 > = |0>, \, exp(f_0 C) | 0 > = exp(f_1 K) | 0 > = exp(f_2 F) | 0 > = |0> \quad (4.21)$$

Thus we now have, writing, $|0,0,0> = |0>$,

$$| S_{m,0}(t) |^2 = | < m_v m_s m_a | exp(f_6 D) exp(f_4 G) | 0,0,0 > |^2 .$$

$$= | < m_v m_s m_a | exp\{b_s^\dagger (f_6 b_v^\dagger + f_4 b_a^\dagger)\} | 0 > |^2 . \quad (4.22)$$

Using the summation expansion of the exponential, we write this as,

$$| < m_v m_s m_a | exp(f_6 D) exp(f_4 G) | 0 > |^2$$

$$= | < m_v m_s m_a | \sum_{n=0} \frac{(b_s^\dagger)^n \{b_v^\dagger f_4 + b_a^\dagger f_6\}^n}{n!} | 0 > |^2. \quad (4.23)$$

Also we write this as a binomial expansion, giving as the double summation,

$$| < m_v m_s m_a | exp(f_6 D) exp(f_4 G) | 0 > |^2 =$$

$$| < m_v m_s m_a | \sum_{n=0}\sum_r \frac{{}^nC_r (b_s^\dagger)^n (b_v^\dagger)^{n-r} (b_a^\dagger)^r (f_4^{n-r} f_6^r)}{n!} | 0 > |^2, \quad (4.24)$$

where $n \geq r$, ${}^nC_r = \dfrac{n!}{(n-r)! r!}$.

Using the orthogonality condition $\delta_{nm} = <m|n>$, we get non-zero terms when $n = m_s$, $n - r = m_v$, $r = m_a$, which gives,

$${}^nC_r = {}^{m_s}C_{m_a} = \frac{m_s!}{(m_s - m_a)! m_a!}, \quad \text{with,} \quad m_s = m_v + m_a . \quad (4.25)$$

Finally the transition probability, $| S_{m,0}(t) |^2$ is,

$$| < m_v m_s m_a | exp(f_6 D) exp(f_4 G) | 0 > |^2 =$$

$$= \frac{m_a!}{m_s! (m_s - m_a)!} | (f_4^{m_v} f_6^{m_a}) |^2 \quad (4.26)$$

and time dependent functions, $f_4(t)$ and, $f_6(t)$ are given by (4.16).

# Conclusion

In this work we have considered three quadratic time dependent Hamiltonians important in the field of quantum optics. Some of the many applications in this field are the use of non - linear crystals to create squeezed states [27,28,29], and entangled Bell states using parametric amplification [30,31]
We have used the group properties of SU(1,1),SU(2) and SU(2,1) to find transition amplitudes for these Hamiltonians. It may also be possible to extend this method to other time dependent quadratic Hamiltonians  Of course, this could have been done by trying to solve Schrödinger's equation for each of the two mode cases see for example[6,7,8] but the time dependent three mode Hamiltonian is more problematic.

In the approach we adopt here we remove the explicit time dependency from the Hamiltonians as a starting point. The methods we employ may not always be applicable when considering other quadratic multi - mode Hamiltonians. The final expression for the transition probabilities (4.26) is rather cumbersome compared to the two mode cases, (1.16) and (2.15),although the point is that the group properties of SU(2,1) can be used to find this transition probability.

Further work using this particular group approach to explicit time dependent quadratic Hamiltonians in quantum optics, as outlined in this paper, will be forthcoming..

# Acknowledgments

Like to thank the referee for bringing to my attention the work of other authors whose work is relavent to this paper and for various corrections to the original script.
Also, Prof Ray Streater, Pamela Jones, Dan Wade of Kings College London, and Nick Herring,Victor. H.Villabobos,Chris Finlay,Margaret Chin and Julia Evans for their constant encouragment.

# Appendix A,Interaction Hamiltonian.

We now prove (2.6) this proof can be found in various text books [21].
Let us write the Schrödinger equation as,

$$ih\frac{\partial \psi_{xs}}{\partial t} = \{H_{xo} + V_x\} \psi_{xs} \tag{A1}$$

then we define,

$$U_{xo} \equiv e^{-iH_{xo}t/h}, \tag{A2a}$$

and

$$\psi_{xs} \equiv U_{xo} \psi_{xI}. \tag{A2b}$$

Then substituting equ(A2b) into equ(A1)gives,

$$ih \frac{\partial U_{xo}}{\partial t}\psi_{xI} + ih U_{xo}\frac{\partial \psi_{xI}}{\partial t} = \{H_{xo} + V_x\}U_{xo}\psi_{xI}, \tag{A3}$$

but from equ(A2a), if $H_{xo}$ is independent of time

$$ih \frac{\partial U_{xo}}{\partial t} = H_{xo} U_{xo} \tag{A4}$$

Substituting equ(A4) into equ(A3) gives,

$$ih\, U_{xo}\frac{\partial \psi_I}{\partial t} = V_x U_{xo}\psi_{xI}, \tag{A5}$$

multiplying equ(A5) on the left by, $U_{xo}^{-1}$ gives

$$ih \frac{\partial \psi_{xI}}{\partial t} = U_{xo}^{-1} V_x U_{xo}\psi_{xI}, \tag{A6}$$

and in equ(A6) we define the interaction Hamiltonian to be (ref 16)

$$H_{xI} \equiv U_{xo}^{-1} V_x U_{xo}. \tag{A7}$$

Finally we can write, using equ(A7), equ(A6) as

$$ih \frac{\partial \psi_{xI}}{\partial t} = H_{xI}\psi_{xI}, \tag{A8}$$

and this has evolution operator, $U_{xI}$.
and this is called the Schrödinger equation for the interaction picture.
Then from (A2b),

$$|\psi_{xs}> \equiv U_{xo} U_{xI}|n> \tag{A9}$$

where $|n>$ is the initial number state.

## Appendix B, SU(3).

We now consider the work of authors [17], in the decomposition of exponential, $e^M$,
where the operator, 'M', usually symbolised by a '3x3' matrix, has SU(3) symmetry.
Using similar notation to that in [17], the authors establish the relation,

$$e^M = exp(\bar{\alpha}\, E)exp(\bar{\beta}\, J)exp(\bar{\gamma}\, B)exp(2C ln\delta)exp(2\{C + 2K - 3F\}ln\varepsilon)exp(\gamma A)exp(\beta G)exp(\alpha D) \tag{B1}$$

where operators, A,B,C etc are the generators of the SU(3) algebra (see below).
'M', in equ(B1), is the matrix,

$$M = \begin{pmatrix} 0 & -z_1 & -z_3 \\ z_1^* & 0 & -z_2 \\ z_3^* & z_2^* & 0 \end{pmatrix}, \tag{B2}$$

with eigenvalue equation,

$$|M - sI| = s^3 + s(|z_1|^2 + |z_2|^2 + |z_3|^2) + z_1 z_2^* z_3^* - z_1^* z_2 z_3 = 0 \qquad (B3)$$

We label it's roots, $s_1$ $s_2$ & $s_3$.

The L.H.S. of equ(B1) is evaluated, using matrix multiplication. We show some of the terms in this matrix,

$$e^M = \begin{pmatrix} \delta\varepsilon & \gamma\delta\varepsilon & \delta\varepsilon(\alpha + \gamma\beta) \\ \overline{\gamma}\delta\varepsilon & \cdot & \cdot \\ (\overline{\alpha} + \overline{\gamma}\overline{\beta})\delta\varepsilon & \cdot & \cdot \end{pmatrix}, \qquad (B4)$$

Using Laplace transforms, the R.H.S. of equ(B1) is also evaluated. We show some of it's terms,

$$e^M = \begin{pmatrix} H - F(|z_1|^2 + |z_3|^2) & -z_1 z_2^* F - z_3 G & z_2 z_3 F - z_1 G \\ z_1^* z_2 F - z_3^* G & \cdot & \cdot \\ z_2^* z_3^* F + z_1^* G & \cdot & \cdot \end{pmatrix} \qquad (B5)$$

where,

$$F = (-1)\frac{(s_2 - s_3)e^{-s_1} + (s_3 - s_1)e^{-s_2} + (s_1 - s_2)e^{-s_3}}{(s_1 - s_2)(s_2 - s_3)(s_3 - s_1)}$$

$$G = \frac{s_1(s_2 - s_3)e^{-s_1} + s_2(s_3 - s_1)e^{-s_2} + s_3(s_1 - s_2)e^{-s_3}}{(s_1 - s_2)(s_2 - s_3)(s_3 - s_1)}$$

$$H = (-1)\frac{s_2 s_3(s_2 - s_3)e^{-s_1} + s_3 s_1(s_3 - s_1)e^{-s_2} + s_1 s_2(s_1 - s_2)e^{-s_3}}{(s_1 - s_2)(s_2 - s_3)(s_3 - s_1)} \qquad (B6)$$

The results equ's (B5) and (B2) are then equated to give $\alpha, \beta, \gamma, \overline{\alpha}, \overline{\beta}, \overline{\gamma}, \delta$, and $\varepsilon$, in terms of the elements of Matrix, 'M'.

We will apply their results to a hypothetical Hamiltonian with SU(3) symmetry.
We will also apply the matrix diagonalising proceedure adopted in the main part of this paper to this Hamiltonian to evaluate the expodential $e^M$.
Let us consider say a hypothetical three mode Hamiltonian, 'H'
with SU(3) symmetry. Let,

$$H = ig_1 h(ab^\dagger - ba^\dagger) + ig_2 h(cb^\dagger - bc^\dagger) + ig_3 h(ca^\dagger - ac^\dagger), \qquad (B7)$$

where in equ(B7), '$H$', is Hermertian i.e., $H = H^\dagger$ and the interaction terms
shown in the brackets are all of the frequency converter type, see section 2.
The number operators in equ(B7) can be labelled as,
A = $ab^\dagger$, B = $ba^\dagger$, and let C = $(-aa^\dagger + bb^\dagger)/2$, \qquad (B8a)
D = $cb^\dagger$, E = $bc^\dagger$, and let F = $(-cc^\dagger + bb^\dagger)/2$, \qquad (B8b)

$G = ca^\dagger$, $J = ac^\dagger$, and let $K = (-cc^\dagger + aa^\dagger)/2$, (B8c)

then we have the commutation relation,
$[A,B] = 2C$, $[D,E] = 2F$, $[G,J] = 2K$. $[A,G] = D$, ect (B9)
We note that these commutation have group the structure,
SU(2) ⊗ SU(2) ⊗ SU(2).

A faithful matrix representation of this group SU(3), under commutation, is,

$$A = \begin{pmatrix} 0 & 1 & 0 \\ 0 & 0 & 0 \\ 0 & 0 & 0 \end{pmatrix}; B = \begin{pmatrix} 0 & 0 & 0 \\ 1 & 0 & 0 \\ 0 & 0 & 0 \end{pmatrix}; C = (1/2)\begin{pmatrix} 1 & 0 & 0 \\ 0 & -1 & 0 \\ 0 & 0 & 0 \end{pmatrix}, D = \begin{pmatrix} 0 & 0 & 1 \\ 0 & 0 & 0 \\ 0 & 0 & 0 \end{pmatrix};$$

$$E = \begin{pmatrix} 0 & 0 & 0 \\ 0 & 0 & 0 \\ 1 & 0 & 0 \end{pmatrix}; F = (1/2)\begin{pmatrix} 1 & 0 & 0 \\ 0 & 0 & 0 \\ 0 & 0 & -1 \end{pmatrix}, G = \begin{pmatrix} 0 & 0 & 0 \\ 0 & 0 & 1 \\ 0 & 0 & 0 \end{pmatrix}; J = \begin{pmatrix} 0 & 0 & 0 \\ 0 & 0 & 0 \\ 0 & 1 & 0 \end{pmatrix};$$

$$K = (1/2)\begin{pmatrix} 0 & 0 & 0 \\ 0 & 1 & 0 \\ 0 & 0 & -1 \end{pmatrix}, \text{ as in equ(B9), } [A,B] = 2C, \text{etc.} \quad (B10)$$

Replacing the number operators with their matrix representation, the Hamiltonian, 'H' can then be written in matrix form.
The evolution operator then becomes,

$$exp(-itH/h) \quad e^M = exp\begin{pmatrix} 0 & g_1 & g_3 \\ -g_1 & 0 & g_2 \\ -g_3 & -g_2 & 0 \end{pmatrix}(t)). \quad (B11)$$

Then using equ's (B5) and (B6), the terms in matrix equ(B11) can be found.
We show one such term i.e.,

$$exp(-itH/h) = \begin{pmatrix} \frac{g_2^2}{(g_3^2+g_2^2+g_1^2)} + \frac{(g_3^2+g_1^2)e^{s_2 t}}{2(g_3^2+g_2^2+g_1^2)} + \frac{(g_3^2+g_1^2)e^{s_3 t}}{2(g_3^2+g_2^2+g_1^2)} & \cdot & \cdot \\ \cdot & \cdot & \cdot \\ \cdot & \cdot & \cdot \end{pmatrix} \quad (B12)$$

Another method for getting this result is acheived by using eigenvectors, as shown earlier in this paper. Briefly, from equ(B11), we diagonalise the matrix
which has eigenvalues $\lambda_i$, to get,

$$\begin{vmatrix} -l_i & g_1 & g_3 \\ -g_1 & -l_i & g_2 \\ -g_3 & -g_2 & -l_i \end{vmatrix} = 0 \qquad \begin{pmatrix} 0 & g_1 & g_3 \\ -g_1 & 0 & g_2 \\ -g_3 & -g_2 & 0 \end{pmatrix} = X \begin{pmatrix} l_1 & 0 & 0 \\ 0 & l_2 & 0 \\ 0 & 0 & l_3 \end{pmatrix} X^{-1} \tag{B13}$$

where matrix ,'X' is constructed from the column eigenvectors ,
and, 'X$^{-1}$', is constructed from row eigenvectors [26], i.e.

$$X^{-1} = \begin{pmatrix} u_1 & v_1 & w_1 \\ u_2 & v_2 & w_2 \\ u_3 & v_3 & w_3 \end{pmatrix} \qquad X = \begin{pmatrix} x_1 & x_2 & x_3 \\ y_1 & y_2 & y_3 \\ z_1 & z_2 & z_3 \end{pmatrix}, \qquad X^{-1}X = I \tag{B14}$$

where, using co - factors of the first row,
$$x_i = (\lambda_i^2 + g_2^2)/N_i, \quad y_i = (-g_1\lambda_i - g_2 g_3)/N_i, \quad z_i = (g_1 g_2 - g_3\lambda_i)/N_i \tag{B15a}$$
and using co - factors of the first column,
$$u_i = (\lambda_i^2 + g_2^2)/N_i, \quad v_i = (g_1\lambda_i - g_2 g_3)/N_i, \quad w_i = (g_1 g_2 + g_3\lambda_i)/N_i. \tag{B15b}$$

We have choosen a suitable normalisation, '$N_i$', so that,
$$x_i u_j + y_i v_j + z_i w_j = \delta_{ij} \tag{B16}$$
Then using equ(B13), we get,

$$\lambda_1 = 0, \; \lambda_2 = -\lambda_3 = i\sqrt{g_3^2 + g_2^2 + g_1^2}$$
$$N_1^2 = g_2^2(g_3^2 + g_2^2 + g_1^2), \; \& \; N_2^2 = N_3^2 = 2(g_3^2 + g_1^2)(g_3^2 + g_2^2 + g_1^2). \tag{B17}$$

These values are then substituted into equ's(B13) & (B11). We show some of the terms i.e. ,

$$\exp\begin{pmatrix} 0 & g_1 & g_3 \\ -g_1 & 0 & g_2 \\ -g_3 & -g_2 & 0 \end{pmatrix}(t) = X \begin{pmatrix} \exp(\lambda_1 t) & 0 & 0 \\ 0 & \exp(\lambda_2 t) & 0 \\ 0 & 0 & \exp(\lambda_3 t) \end{pmatrix} X^{-1} =$$

$$\begin{pmatrix} x_1 u_1 e^{\lambda_1 t} + x_2 u_2 e^{\lambda_2 t} + x_3 u_3 e^{\lambda_3 t} & x_1 v_1 e^{\lambda_1 t} + x_2 v_2 e^{\lambda_2 t} + x_3 v_3 e^{\lambda_3 t} & x_1 w_1 e^{\lambda_1 t} + x_2 w_2 e^{\lambda_2 t} + x_3 w_3 e^{\lambda_3 t} \\ y_1 u_1 e^{\lambda_1 t} + y_2 u_2 e^{\lambda_2 t} + y_3 u_3 e^{\lambda_3 t} & y_1 v_1 e^{\lambda_1 t} + y_2 v_2 e^{\lambda_2 t} + y_3 v_3 e^{\lambda_3 t} & \cdot \\ z_1 u_1 e^{\lambda_1 t} + z_2 u_2 e^{\lambda_2 t} + z_3 u_3 e^{\lambda_3 t} & \cdot & \cdot \end{pmatrix}.$$

(B16)

Together with equ(B15) we find the terms in matrix equ(B16). We show one such term,

$$x_1 u_1 e^{\lambda_1 t} + x_2 u_2 e^{\lambda_2 t} + x_3 u_3 e^{\lambda_3 t} = \frac{g_2^2}{(g_3^2 + g_2^2 + g_1^2)} + \frac{(g_3^2 + g_1^2)e^{\lambda_2 t}}{2(g_3^2 + g_2^2 + g_1^2)} + \frac{(g_3^2 + g_1^2)e^{\lambda_3 t}}{2(g_3^2 + g_2^2 + g_1^2)}. \tag{B17}$$